\documentclass[prb,aps,showpacs]{revtex4}

\usepackage{graphicx}
\usepackage{ifthen}
\usepackage{ifpdf}

\usepackage{latexsym}
\usepackage{amssymb}
\usepackage{bm}
\newcommand{\half}{\mbox{\small $\frac{1}{2}$}}

\newcommand{\eexp}{\mbox{e}^}
\newcommand{\bra}{\left\langle}
\newcommand{\ket}{\right\rangle}

\newcommand{\beq}[1]{\begin{eqnarray}\ifthenelse{#1=-1}{\nonumber}
{\ifthenelse{#1=0}{}{\label{e#1}}}}
\newcommand{\eeq}{\end{eqnarray}}
\newcommand{\hide}[1]{}
\newcommand{\RR}{\bf R}
\newcommand{\qq}{\bf q}

\begin{document}

\title{The Aharonov-Bohm effect in presence of dissipative environments}
\author{Baruch Horovitz{$^1$} and Pierre Le Doussal{$^2$} }
\affiliation{{$^1$} Department of Physics, Ben Gurion university,
Beer Sheva 84105 Israel} \affiliation{{$^2$}CNRS-Laboratoire de
Physique Th{\'e}orique de l'Ecole Normale Sup{\'e}rieure, 24 rue
Lhomond,75231 Cedex 05, Paris France.}

\begin{abstract}
We study a particle on a ring in presence of various dissipative
environments. We develop and solve a variational scheme assuming
low frequency dominance. Our solution produces a renormalization group (RG)
 transformation to all orders in the inverse dissipation strength,
 and in particular reproduces known two loop results. Our RG leads
 to a weak dissipation parameter, for which a weak coupling expansion for the position correlation function shows a $1/\tau^2$ decay in imaginary time.
\end{abstract}

\maketitle

\section{Introduction}
The problem of interference and dephasing in presence of
dissipative environments is of significance for a variety of
experimental systems and a fundamental theoretical issue. The
experimental systems include mesoscopic rings embedded on various
surfaces where Aharonov-Bohm (AB) oscillations can be measured
\cite{web,jariwala}, and the related problem of decoherence at low
temperatures \cite{mohanty}. A different type of experimental
systems are cold atom traps created by atom chips
\cite{harber,jones,lin}. The atom chip that produces a magnetic or
electric trap for the cold atoms necessarily also produces noise.
Our problem is then relevant for evaluating the interference
amplitude of the cold atoms in presence of such noise.

As an efficient tool for monitoring the effect of the environment
we follow a suggestion by Guinea \cite{guinea} to find the AB
oscillation amplitude as function of the radius R of the ring; for
free particles of mass M this amplitude is the mean level spacing
$\sim 1/MR^2$. Two types of environments were suggested to lead to
an anomalous suppression, i.e. a stronger decrease of the
oscillation amplitude than $1/R^2$:  a
Caldeira Legget (CL) bath as well as a charge - metal (CM)  system, i.e. a charge on the ring interacting with a dirty metal
environment. The CL system is of further interest since it can be mapped to
 the Coulomb blockade problem \cite{hofstetter,herrero} as well as to quantum dots at a distance
from metallic gates \cite{guinea2}. The Coulomb box problem is of further recent interest in view of data on the quantization of the charge relaxation resistance \cite{gabelli,feve} and related theoretical developments \cite{mora, burmistrov1,burmistrov2}.

The CL system has been
extensively investigated by instanton methods
\cite{panyukov,wang}, by RG methods \cite{guinea,hofstetter}, by a boundary field theory \cite{lukyanov1} and
by Monte Carlo (MC) methods \cite{hofstetter,herrero,lukyanov2}. All methods
show that the effective mass, defined as $B/R^2$, of the particle
increases exponentially with the dissipation strength $\alpha$,
i.e. $B\sim \alpha^{\mu}e^{\pi^2\alpha}$, with differences in the
exponent $\mu$. In 2nd order renormalization group (RG) $\mu=-1$
\cite{hofstetter} while instanton methods give either
\cite{panyukov} $\mu=-2$ or \cite{wang} $\mu=-3$; the boundary field theory with MC gives $\mu=-2$. A variational
approach \cite{brown} indicated a nonperturbative regime at strong
$\alpha$. Since $\alpha=\gamma R^2$ where $\gamma$ is a friction
coefficient, a length scale $\pi/\sqrt{\gamma}$ is identified
\cite{guinea}; this scale is a candidate for a dephasing length.

The CM system was investigated by RG methods \cite{guinea}
finding $B\sim R^{2+\mu '}$ with $\mu'\lesssim 1$ nonuniversal,
while MC data \cite{golubev} shows $\mu ' \approx 1.8$.
Further MC simulations show that in fact $\mu'=0$, at least for weak coupling \cite{kagalovsky}.
We study also a dipole-metal (DM) system, i.e. an electric dipole on a ring coupled to a dirty metal environment. This system can be realized by experiments
 on cold Rydberg atoms \cite{hyafil}.

In the present work, extending our previous report \cite{bh} we solve these systems by a variational
method, assuming low frequency dominance. We find that the variational method defines
an RG scheme to all orders, reproducing a known RG equation
\cite{hofstetter} to two loops in the CL system. In the CM and DM systems, for either a charge or a dipole, we find that the effective mass
remains $B/R^2\sim R^0$ for large $R$, as for
free particles. Our RG leads
 to a weak coupling dissipation parameter. The resulting action yields a weak coupling expansion for the position correlation function, showing a $1/\tau^2$ decay in imaginary time.
 This decay is generic to all finite $R$ systems and indicates dephasing of an excited state. In the limit $R\rightarrow\infty$ the correlation probes degenarate states, however, the position correlation function does not decay in this limit, i.e. no dephasing.

In section II we present the models. In section III we define our
variational method and show that the effective mass $B$ of the
$m=0$ sector determines the curvature $\partial^2E_0/\partial
\phi_x^2|_0$ where $E_0$ is the ground state energy and $\phi_x$ is the flux
through the ring; this
curvature is a measure of the Aharonov Bohm oscillation amplitude.
In section IV we simplify the variational equation by assuming low
frequency dominance, or equivalently logarithmic dominance. In
section V we show that this method is equivalent to an RG scheme, and in particular
reproduces the known RG equation to 2nd order in the CL system. In
general the variational equation contains terms to all orders and
is therefore expected to be superior to a 2nd order RG expansion. In section VI we present
explicit solutions for the CL and DM systems, as well as for a general case.
Finally in section VII we study the weak coupling expansion showing a $1/\tau^2$ decay for the position correlation function.

\section{The Model}

In this section we derive the effective action in presence of a
dissipative environment in terms of the the angle $\theta_m(\tau)$ where $\tau$ is an imaginary time. The index $m$ specifies the winding number so that
\begin{equation}\label{winding}
\theta_m(\tau)=\theta(\tau)+2\pi m\tau/\beta
\end{equation}
where $\theta(0)=\theta(\beta)$ has periodic boundary condition
and $\beta$ is the inverse temperature ($\beta\rightarrow\infty$
below). In presence of an external flux $\phi_x$ the partition sum
has the form
\begin{equation}\label{Z}
Z=\sum_m \int {\cal D}\theta e^{2\pi
im\phi_x-S_1\{\theta_m\}-S_{int}\{\theta_m\}}
\end{equation}
As shown by Guinea \cite{guinea}, the form of such an action in presence of a general dissipative bath the effective
action can be written in terms of a kernel
$K[\theta(\tau)-\theta(\tau')]$ that is periodic and allows in
general a Fourier expansion
\begin{eqnarray}\label{S}
S_1\{\theta\}&=&\int_0^{\beta}d\tau
\frac{MR^2}{2}(\frac{\partial \theta}{\partial
\tau})^2\nonumber\\
S_{int}\{\theta\}&=&\alpha\int_0^{\beta}\int_0^{\beta}d\tau
d\tau ' \frac{\pi ^2 T^2 K[\theta(\tau)-\theta(\tau
')]}{\sin ^2[\pi T(\tau -\tau ')]}\nonumber\\
&=&\alpha\sum_n a_n\int_0^{\beta}\int_0^{\beta}d\tau d\tau '
\frac{\pi ^2 T^2\sin^2 \{n[\theta(\tau)-\theta(\tau
')]/2\}}{\sin ^2[\pi T(\tau -\tau ')]}
\end{eqnarray}
and $a_n$ depends on the type of bath. At $\tau\rightarrow\tau'$
(or at high frequencies $\omega$) one can expand the $\sin^2(...)$
in (\ref{S}) and then $S_{int}\rightarrow \alpha\sum_n a_nn^2 \int
d\omega |\omega||\theta_m (\omega)|^2$, identifying a dissipative
system.

We consider now 3 types of environments and identify the
coefficients $a_n$. First is the Caldeira Legget (CL) environment. It has harmonic
oscillators coupled linearly to the particle's coordinate. The
effective action is well known \cite{CL} for nonconfined coordinate
$\RR(\tau)$
\beq{01} S_{int}^{CL}=\gamma\int\int d\tau
d\tau'\frac{\pi^2T^2[\RR(\tau)-\RR(\tau')]^2}{\sin^2[\pi
T(\tau-\tau')]}\}\eeq
where $\gamma$ is the dissipation parameter. When the particle is
confined to a ring
$\RR(\tau)=R[\cos\theta(\tau),\sin\theta(\tau)]$ the action
becomes of the form of Eq. (\ref{S}) with a single coefficient
$a_1=1$ and $\alpha=\gamma R^2$.

Consider next the charge-metal (CM) environment. It consists of a dirty metal that is
characterized by its conductivity $\sigma$ and diffusion constant
$D$. The particle on the ring has a charge $e$ and responds to the
Coulomb potential of the metal $V(\RR(\tau),\tau)$. The metal is
assumed to be a Gaussian environment, so that the interaction term
(in imaginary time) of the partition sum can be averaged to obtain
\cite{golubev}
\beq{02} Z_{int}=\bra
\eexp{-i\int_0^{\beta}V(\RR(\tau),\tau)d\tau}\ket\equiv
\eexp{-S_{int}}\eeq
 and with $\int_q\equiv \int d^3q/(2\pi)^3$
 \beq{03} S_{int}=\half e^2 \bra V(\RR(\tau),\tau)
V(\RR(\tau'),\tau')\ket =\half e^2
T\sum_n\int_q\frac{4\pi}{q^2\epsilon(i|\omega_n|,q)}
\eexp{i\qq\cdot(\RR(\tau)-\RR(\tau'))-i\omega_n(\tau-\tau')}\eeq
where the propagator of the scalar potential \cite{AGD} is given
in terms of the dielectric function $\epsilon(i|\omega_n|,q)$ with
$\omega_n=2\pi n T$ the Matsubara frequencies. At low frequencies
and momenta
$\epsilon(\omega,q)=1+\frac{4\pi\sigma}{-i\omega+Dq^2}$, valid at
$q<1/\ell$ where $\ell$ is the mean free path. Hence
$1/\epsilon(i|\omega_n|,q)\approx (|\omega_n|+Dq^2)/4\pi\sigma$;
the $Dq^2$ term yields an $\RR(\tau)$ independent constant while
\beq{04} \sum_n|\omega_n|\eexp{-i\omega_n\tau}=\frac{-\pi
T^2}{\sin^2[\pi
 T\tau]}\eeq
hence  with $k_F$ the Fermi wavevector and $r=R/\ell$ the charge coupled to a dirty metal has
\beq{05}
\alpha&=&\frac{3}{8 k_F^2\ell^2}\nonumber\\
K(z)&=&1-[4r^2\sin^2(z/2)+1]^{-1/2}\eeq
For $r\gg 1$, $a_n\approx \frac{2}{\pi r}\ln
(r/n)$ for $1<n\lesssim r$ and $a_n\approx 0$ for $n\gtrsim r$.
This model reduces to the CL one
 at $r\ll 1$, where $a_1$ single
$a_n$ survives.

A 3rd realization of the action corresponds to the dipole-metal (DM) environment. Consider a particle with an
electric dipole, whose direction is perpendicular to the ring, interacting with a metal.
For the electric field
$E_z=\partial_zV-\frac{1}{c}\partial_tA_z$, the $A_z$ propagator
involves \cite{AGD}
$[\epsilon(i|\omega_n|,q)\omega_n^2+Dq^2]^{-1}$, which for $q\neq
0$ can be expanded in $\omega_n^2$, hence it has no dissipative
term $\sim|\omega_n|$; we keep then just the $\partial_z V$
term. The interaction with the fluctuating electric
field ${\bf E}({\bf r},\tau)$ is $p\int_0^{\beta}E_z({\bf
R}(\tau),\tau)d\tau$. A Gaussian average on the metallic environment then yields
\begin{eqnarray}\label{dipole}
S_{int}=\half p^2\int_{\tau}\int_{\tau'} \bra \partial_z V(\RR(\tau),\tau)
\partial_z V(\RR(\tau'),\tau')\ket =\half p^2
T\sum_n\int_q\frac{4\pi q_z^2}{q^2\epsilon(i|\omega_n|,q)}
\eexp{i\qq\cdot(\RR(\tau)-\RR(\tau'))-i\omega_n(\tau-\tau')}
\end{eqnarray}
Therefore
\beq{010}
\alpha&=&\frac{3}{8k_F^2\ell^2}\frac{p^2}{e^2\ell^2}\nonumber\\
K(z)&=&1-(4r^2\sin^2\frac{z}{2}+1)^{-3/2}\eeq
Hence, for large $r$, $a_n\sim\frac{1}{r}(1-\frac{n^2}{r^2})$
for $n\lesssim r$ and $a_n\approx 0$ otherwise. Finally we note that a topological flux $\phi_x$ can be realized for an electric dipole \cite{spavieri}.

\section{Variational Method}

The partition sum can be rewritten by using the Poisson sum
$\sum_m g(m)=\sum_K\int g(\phi)\exp (2\pi i K\phi)d\phi$ so that
\begin{eqnarray}\label{Z1}
Z&=&\sum_K\int_{-\infty}^{\infty}d\phi\int {\cal D}\theta e^{2\pi
i\phi(K+\phi_x)-\frac{2\pi^2MR^2\phi^2}{\beta}
-S_1\{\theta(\tau)\}-S_{int}\{\theta(\tau)+2\pi
\phi\tau/\beta\}}\nonumber\\
&=&\sum_K\int_{-\infty}^{\infty}d\phi e^{2\pi
i\phi(K+\phi_x)-\frac{2\pi^2MR^2\phi^2}{\beta}} \, Z_{\phi}
\end{eqnarray}

The variational method for $Z_{\phi}$ finds the best Gaussian
approximation, i.e.
\begin{equation}
S_0=\frac{1}{2\beta}\sum_{\omega_n}G^{-1}(\omega_n)|\theta(\omega_n)|^2
\end{equation}
so that the free energy $\cal F$ in $Z_{\phi}=\eexp{-\beta{\cal F}}$ has the variational form
\begin{equation}\label{Fvar}
\beta F_{var}=\beta F_0+\langle
S-S_0\rangle_0=\frac{1}{2\beta}\sum_{\omega_n}\{-\ln
G(\omega_n)+[MR^2\omega_n^2-G^{-1}(\omega_n)]G(\omega_n)\}+\langle
S_{int}\rangle_0
\end{equation}
where $\langle ... \rangle_0$ is an average with respect to
$\exp(-S_0)$ and $F_0$ is the free energy corresponding to $S_0$.
Since
\begin{eqnarray}
&&2\langle \sin^2 \{n[\theta(\tau)-\theta(\tau
')+2\pi\phi(\tau-\tau')/\beta]/2\}\rangle_0\nonumber\\
&&=1-\cos[2\pi n\phi(\tau-\tau')/\beta]\exp\{-n^2\langle
[(\theta(\tau)-\theta(\tau
')^2]\rangle_0/2\}\nonumber\\
&&=1-\cos[2\pi
n\phi(\tau-\tau')/\beta]\exp\{-n^2T\sum_{\omega_n}G(\omega_n)[1-\cos
(\omega_n \tau)]\}
\end{eqnarray}
the interaction term becomes
\begin{equation}\label{Sint}
\langle S_{int}\rangle_0=\beta \alpha\sum_na_n \int_{0}^{\beta}
\frac{d\tau}{2\tau^2}\{1-\cos (2\pi
n\phi\tau/\beta)e^{-n^2T\sum_{\omega}G(\omega)[1-\cos
(\omega\tau)]}\} \,.
\end{equation}
The variational equation $\delta F_{var}/\delta G(\omega_n)=0$ is
then
\begin{equation}\label{G}
G^{-1}(\omega)=MR^2\omega^2+2\alpha\sum_na_n
n^2\int_{0}^{\beta}d\tau \frac{1-cos(\omega\tau)}{\tau^2}\cos
(2\pi
n\phi\tau/\beta)e^{-n^2\int(d\omega_1/2\pi)G(\omega_1)[1-\cos
(\omega_1\tau)]}
\end{equation}
When the limit $\beta\rightarrow \infty$ is taken a cutoff
$\omega_c$ may be introduced to control the short time behavior so
that the $\tau$ integral becomes $\int_{1/\omega_c}^{\infty}$.
This cutoff represents a high frequency limit of the bath degrees
of freedom. Alternatively, the mass term serves also as a cutoff
since it leads to convergence of the $d\omega_1$ integral in the
exponent of (\ref{G}).

In the following we will study the variational equation with
$\phi=0$. To justify this, we show now that the effective mass $B$
of the $\phi=0$ system is indeed what is needed to find the
Aharonov-Bohm oscillation amplitude at $T\rightarrow 0$. The
effective mass is defined by $G^{-1}(\omega)=B\omega^2$ in the
limit $\omega\rightarrow 0$ and is identified from Eq. (\ref{G})
at $\beta\rightarrow \infty$ as
\begin{equation}\label{B0}
B=MR^2+\frac{1}{2}\alpha\sum_na_nn^2\int_0^{\infty} d\tau
e^{-n^2\int(d\omega/2\pi)G(\omega)[1-\cos (\omega\tau)]}
\end{equation}
The form (\ref{Z1}) implies that the fluctuations
$<\phi^2>\sim\beta$, hence the factor $\cos (2\pi
n\phi\tau/\beta)\rightarrow 1+O(1/\beta)$ in Eq. (\ref{G}) and the
effective mass $B$ is $\phi$ independent. It is also necessary to
check that the $\tau$ integrals converge: indeed at
$\tau\rightarrow \infty$
\[\int_0^{\infty}d\omega G(\omega)(1-\cos(\omega\tau)\approx
\tau^2\int_0^{1/\tau}\frac{d\omega}{2B}+\int_{1/\tau}^{\infty}d\omega
G(\omega)\sim \tau/B \] hence a factor $e^{-n^2\tau/B}$ assures
the convergence of the $\tau$ integrals.

The Aharonov-Bohm oscillation amplitude is usually measured
\cite{hofstetter,herrero} by the curvature of the free energy at $\phi_x=0$;  since at
$\phi=0$ we have $\partial G(\omega)/\partial \phi=0$ (from parity in $\phi$, see (\ref{G}),
and from analyticity in $\phi$) and
$\partial F_{var}/\partial G=0$ (the variational condition) we
obtain from Eqs. (\ref{Fvar},\ref{Sint},\ref{B0})
\begin{equation}
\frac{\partial^2 \beta F_{var}}{\partial
\phi^2}|_0=\frac{\partial^2 \beta \langle
S_{int}\rangle_0}{\partial
\phi^2}|_0=\frac{4\pi^2}{\beta}(B-MR^2)\,.
\end{equation}
The effect of $Z_{\phi}$ in the partition sum Eq. (\ref{Z1}) is
therefore to replace the factor $2\pi^2MR^2\phi^2/\beta$ by
$2\pi^2B\phi^2/\beta$, i.e. the response to an external flux is
that of a free particle with a mass renormalized to $B$. Higher order terms
produce only subdominant behavior in $1/\beta$, e.g. one expects a $\phi^4/\beta^3$ term.
Our task
is therefore to study the $\phi=0$ system and find this
renormalized mass.

\section{variational equation}
Before studying the full equation, it is instructive to study its perturbative regime.
The lowest order is obtained by neglecting the exponent in Eq. (\ref{G}), leading to
\begin{equation}\label{Gpert}
G^{-1}(\omega)=MR^2 \omega^2+\pi \omega \alpha\sum_na_n n^2 \qquad \omega\lesssim \omega_c
\end{equation}
This identifies the cutoff $\omega_c$ below which dissipative term dominates,
\begin{equation}\label{wc}
\omega_c=\frac{\pi\alpha\sum_na_nn^2}{MR^2}
\end{equation}
Consider next $\omega\ll \omega_c$ but still $\ln(\omega_c/\omega)\lesssim 1$. The next order in perturbation is obtained by using Eq. (\ref{Gpert}) in the exponent in Eq. (\ref{G}) and expanding this exponent,
\begin{equation}\label{Gpert2}
G^{-1}(\omega)=\pi \omega \alpha\sum_na_n
n^2[1-\frac{n^2}{\pi^2\alpha\sum_ma_m m^2}\ln
\frac{\omega_c}{\omega}]=\pi\omega\alpha\sum_na_nn^2[1-\frac{1}{\alpha\kappa}\ln
\frac{\omega_c}{\omega}] \qquad \ln(\omega_c/\omega)\lesssim 1
\end{equation}
 where the mass term is ignored for $\omega< \omega_c$ and $\kappa$ is a geometric parameter defined by
\begin{equation}\label{alphac}
\kappa= \frac{\pi^2(\sum_na_nn^2)^2}{\sum_na_nn^4}
\end{equation}
The sums in (\ref{alphac}) can be evaluated for each model from the 2nd and 4th derivatives of $K(z)$ at $z=0$, leading to
\begin{eqnarray}
\kappa &=\pi^2 \qquad \qquad \qquad  & {\text CL }\nonumber\\
&=\frac{2\pi^2r^4}{r^2+9r^4} \qquad \qquad &{\text CM }\nonumber\\
&=\frac{6\pi^2 r^4}{r^2+15r^4}  \qquad \qquad &{\text DM }
\end{eqnarray}

A significant perturbative regime is possible for $\alpha\kappa\gg 1$.
This strong dissipation condition can apply to the  CL model if $R$ is large, though one needs to make sure that the CL model is still valid in that case. For the CM or DM models $\kappa$ is bounded by a number $\sim 1$ so that $\alpha\gg 1$ is needed. For usual dirty metals $k_F\ell\gtrsim 1$ so that for
charge coupling with Eq. (\ref{e05}) the condition is not satisfied, unless the
particle on the ring has a charge $e^*\gg e$. On the other hand, the dipole
case may have a large $\alpha$ in Eq. (\ref{e010}) for large dipoles, e.g. in Rydberg atoms.
In the following we
use the form (\ref{Gpert2}) as a boundary condition for the full variational solution.

We proceed now to variational equation, that includes the significant range of $\omega\ll \omega_c$.
It is convenient to study a derivative of Eq. (\ref{G})
\begin{equation}\label{G'}
\frac{d}{d\omega}G^{-1}(\omega)=2\alpha\sum_na_n
n^2 \int_{0}^{\infty}d\tau
\frac{\sin(\omega\tau)}{\tau}e^{-n^2\int(d\omega_1/2\pi)G(\omega_1)[1-\cos
(\omega_1\tau)]}
\end{equation}
The bare mass $M$ serves to define $\omega_c$ and then the $MR^2\omega^2$ term in Eq. (\ref{G}) is neglected at $\omega<\omega_c$.
If $\omega$ is sufficiently small then $\sin(\omega\tau)$ can be
expanded leading to a $\sim \omega$ term. We therefore assume the
form
\begin{eqnarray}\label{forms}
G^{-1}&=&f(\omega) \qquad   \omega_0<\omega< \omega_c
\nonumber\\
G^{-1}&=& B\omega^2    \qquad  \omega<\omega_0
\end{eqnarray}
The solution for $f(\omega)$ needs to satisfy boundary conditions, whose $\alpha$ dependence is determined by the perturabtive expansion Eq. (\ref{Gpert2}),
\begin{eqnarray}\label{bc}
f(\omega_c)&=\pi \omega_c \alpha\sum_na_n n^2 \qquad \qquad &\nonumber\\
f'(\omega_c)&=\pi  \alpha\sum_na_n n^2\cdot \eta(\alpha) \qquad  &\eta(\alpha)=1+\frac{1}{\alpha\kappa}+O(1/\alpha^2)\nonumber\\
f''(\omega_c)&=\pi  \sum_na_n n^2\cdot\frac{C(\alpha)}{\omega_c} \qquad &C(\alpha)=\frac{1}{\kappa}+O(1/\alpha)
\end{eqnarray}

We proceed to simplify Eq. (\ref{G'}). For $\omega>\omega_0$ the oscillating $\sin (\omega \tau)$ in Eq.
(\ref{G'}) leads to a cutoff $\tau<1/\eta_1 \omega$,  to be determined
by matching to the perturbative regime. Hence
\begin{equation}\label{f'}
f'(\omega)=2\omega\alpha\sum_na_n n^2
\int_{0}^{1/\eta_1\omega}d\tau
e^{-n^2\int_0^{\omega_c}(d\omega_1/2\pi)G(\omega_1)[1-\cos
(\omega_1\tau)]}
\end{equation}
The range $\int_{\omega_c}^{\infty}$ involves $G^{-1}(\omega)=MR^2\omega^2$
and contributes $\sim 1/(MR^2\omega_c)=[\pi\alpha\sum_na_nn^2]^{-1}$ which
is neglected for $\alpha\gg 1$. The $\tau$ integration is dominated by
$\tau\approx 1/\eta_1 \omega$, hence $1-\cos\omega_1\tau\approx
1-\cos(\omega_1/\eta_1\omega)$ is replaced by
$\omega_1^2/2\omega^2\eta_1^2$ for $\omega_1<\omega$ and by $1$
for $\omega_1>\omega$. This rough separation is to be justified by
our main assumption that
$\int_{\omega}^{\omega_c}d\omega_1/f(\omega_1)$ dominates this
integral due to the low frequency decrease of $f(\omega_1)$. The
terms from $\omega_0<\omega_1<\omega$, as well as those from
$\omega_1<\omega_0$, can be neglected if
\begin{equation}\label{cond1}
\frac{1}{B\omega_0},\,\, \frac{1}{\omega^2}\int_{\omega_0}^{\omega}
\frac{\omega_1^2d\omega_1}{f(\omega_1)}
\ll \int_{\omega}^{\omega_c}\frac{d\omega_1}{f(\omega_1)} \qquad
\mbox{condition (i)}\,.
\end{equation}
Note that the 2nd term on the left near $\omega_0$ is $\sim 1/B\omega_0$, while near $\omega_c$ it is
$\sim 1/\alpha$ and negligible for large $\alpha$.
We are interested in nonperturbative contributions, i.e. the range $\ln \omega_c/\omega\ll 1$ and in particular at $\omega=\omega_0$.
In terms of $\omega_2=1/\tau$ we obtain
\begin{equation}\label{f2}
f'(\omega)=2\omega\alpha\sum_na_n
n^2\int_{\eta_1\omega}^{\infty}\frac{d\omega_2}{\omega_2^2}
e^{-n^2\int_{\eta_2\omega_2}^{\omega_c}d\omega_1/\pi f(\omega_1)}
\end{equation}
where as above, the precise location of the $\eta_2\omega$ cutoff
should not be significant. The $\omega_2$ integration is dominated
by its lower cutoff $\eta_1 \omega$ so we expect that the exponent
can be taken out of the integration with the replacement
$\eta_2\omega_2\rightarrow \eta_1\eta_2 \omega$. More precisely,
taking a derivative of (\ref{f2}) leads to
\begin{equation}\label{f4}
f'(\omega)=\pi {\tilde\eta}(\alpha)\alpha \sum_na_n n^2
e^{-n^2\int_{\omega}^{\omega_c}d\omega_1/\pi f(\omega_1)}+\omega f''(\omega)
\end{equation}
and $\eta_1=2/(\pi {\tilde\eta})$ and $\eta_1\eta_2=1$ are chosen. The coefficient ${\tilde\eta}(\alpha)$ is to be determined by the boundary conditions (\ref{bc}).

To further simply the equation we assume now
\begin{equation}\label{cond2}
f''(\omega) \ll
\frac{f'(\omega)}{\omega} \qquad \mbox{condition (ii)}\,.
\end{equation}
leading to our main equation for $f(\omega)$,
\begin{equation}\label{f5}
f'(\omega)=\pi \eta(\alpha) \alpha\sum_na_n n^2
e^{-n^2\int_{\omega}^{\omega_c}d\omega_1/\pi f(\omega_1)}
\end{equation}
The coefficient here is $\eta(\alpha)$, consistent with (\ref{bc}).
Below we actually find that condition (ii) is not always satisfied, and then we return to solve Eq. (\ref{f4}) instead of (\ref{f5}).

Finally, consider $\omega<\omega_0$. Eq. (\ref{G'}) has then on the left
$\frac{d}{d\omega}G^{-1}(\omega)=2B\omega$ while on the right it has the requested $\sim\omega$ form, except for a term
where
\begin{eqnarray}
I_1&=&2\omega\alpha\sum_na_n n^2
\int_{1/\omega_0}^{1/\eta_1\omega}d\tau\exp[-\frac{n^2}{\pi}(\int_0^{\omega_0}\frac{1-\cos
\omega_1\tau}{B\omega_1^2}d\omega_1+\int_{\omega_0}^{\omega_c}\frac{d\omega_1}{f(\omega_1)})]
\end{eqnarray}
 Since $\tau>1/\omega_0$ dominates, $\int_0^{\omega_0}\omega_1^{-2}(1-\cos
\omega_1\tau)d\omega_1\approx \pi \tau/2$, hence
\begin{equation}\label{I1}
I_1=4\omega
B\alpha\sum_na_n(e^{-n^2/2B\omega_0}-e^{-n^2/2B\eta_1\omega})
e^{-n^2\int_{\omega_0}^{\omega_c}d\omega_1/\pi f(\omega_1)}
\end{equation}
The essential singularity in $\omega$ is negligible for $\omega<\omega_0$ when
\begin{equation}\label{cond3}
B\omega_0\gtrsim 1   \qquad \mbox{condition (iii)}
\end{equation}
The remaining term at $\omega<\omega_0$ identifies B and leads to a
matching condition of the form  $f'(\omega_0)=\eta'
B\omega_0$. Continuity of derivatives yeilds $\eta'=2$, though we
expect that the precise value of $\eta'$ will not be significant.

This completes the derivation of the equations for $B$ and
$f(\omega)$. Eq. (\ref{f4}) or (\ref{f5}) are to be solved with the boundary
conditions in (\ref{bc}) (in case of (\ref{f5}) only the first two conditions are needed).  Furthermore, the matching conditions at $\omega_0$ are
\begin{eqnarray}\label{allbc}
f(\omega_0)&=&B\omega_0^2\nonumber\\
f'(\omega_0)&=&\eta' B\omega_0 \qquad \eta'\approx 2
\end{eqnarray}

\section{RG procedure}

We present here an approximate solution of the variational equations by an RG method, which in some case (the CL case, see below) should be very close to exact. The idea is that an $\omega<\omega_c$ can serve as a new cutoff provided that the coupling $\alpha$ is renormalized into ${\bar \alpha}(\omega)$. The boundary conditions (\ref{bc}) become therefore
\begin{eqnarray}\label{rg1}
f(\omega)&=&\pi \omega {\bar \alpha}(\omega)\sum_na_n n^2 \nonumber\\
f'(\omega)&=&\pi {\bar \alpha}(\omega)\sum_n a_n n^2\cdot \eta[{\bar \alpha}(\omega)] \nonumber\\
f''(\omega)&=&\pi  \sum_n a_n n^2\cdot\frac{C[{\bar \alpha}(\omega)]}{\omega}
\end{eqnarray}
The number of needed equations depends on the order of the differential equation for $f(\omega)$, e.g. for Eq. (\ref{f5}) only the first two equations in (\ref{rg1}) are needed. The functions $\eta(\alpha), C(\alpha)$ are known as an expansion in $1/\alpha$. As we find below, these functions can be determined explicitly by the variational equations.

Taking a derivative of the 1st equation in (\ref{rg1}) yields a recursion relation for ${\bar \alpha}(\omega)$,
\begin{equation}\label{rg2}
\omega\frac{d{\bar \alpha}(\omega)}{d\omega}={\bar \alpha}(\omega)[\eta({\bar \alpha}(\omega))-1]
\end{equation}
Hence the boundary condition function $\eta(\alpha)$ determines the flow of the renormalized ${\bar \alpha}(\omega)$, i.e. it generates the RG flow to all orders for which $\eta(\alpha)$ is known. In particular the flow terminates when $\eta(\alpha_c)=1$, i.e. $\alpha_c$ is a fixed point.

Before proceeding to solve for $\eta(\alpha)$, we show that the RG is equivalent to a solution of the form $f(\omega)=\omega g(K(\alpha)\omega)$, so that all the $\alpha$ dependence is included in the function $K(\alpha)$, i.e. the function $g(x)$ itself is $\alpha$ independent. This property is exact for our variational equation for the CL system (see below and Appendix \ref{app:log}). For other systems the scaling function needs to be identified separately, as e.g. done in section VIC for the CM system.

The first boundary condition from (\ref{bc}) is $g(K(\alpha)\omega_c)=\pi\alpha  \sum_n a_n n^2$, hence $K'(\alpha)\omega_cg'(K(\alpha)\omega_c)=\pi  \sum_n a_n n^2$. The second boundary condition is then
\begin{eqnarray}\label{rg3}
f'(\omega_c)&=&g(K(\alpha)\omega_c)+K(\alpha)\omega_cg'(K(\alpha)\omega_c)=
(\alpha+ \frac{K(\alpha)}{K'(\alpha)}) \pi  \sum_n a_n n^2 =\pi\alpha\eta(\alpha) \sum_n a_n n^2 \nonumber\\
\Rightarrow && \eta(\alpha)=1+\frac{K(\alpha)}{\alpha K'(\alpha)}
\end{eqnarray}

In $g(K(\alpha)\omega)$ one can vary either $\alpha$ or $\omega$ with identical effects if
$K(\alpha)\omega=K({\bar \alpha}(\omega))\omega_c$ (see also Appendix B) which by $d/d\omega$ yields
\begin{equation}\label{rg4}
K'[{\bar \alpha}(\omega)]\frac{d{\bar \alpha}(\omega)}{d\omega}\omega_c=K(\alpha)=K[{\bar \alpha}(\omega)]\frac{\omega_c}{\omega}
\end{equation}
and with (\ref{rg3}) the flow (\ref{rg2}) is reproduced. We note also that the scaling functions $\eta(\alpha),\,C(\alpha)$ (\ref{rg1}) can also be determined by rewriting the differential equation for $g(x)$ in terms of $y(g)=x(g)g'(x(g))$ and its derivatives. This is possible under fairly general conditions, e.g. that $g(x)$ is monotonic and that the differential equation for $g(x)$ is homogenous (i.e. contains only $x^ng^{(n)}(x)$ terms).

\begin{figure}[b]
\begin{center}
\includegraphics[scale=1.0]{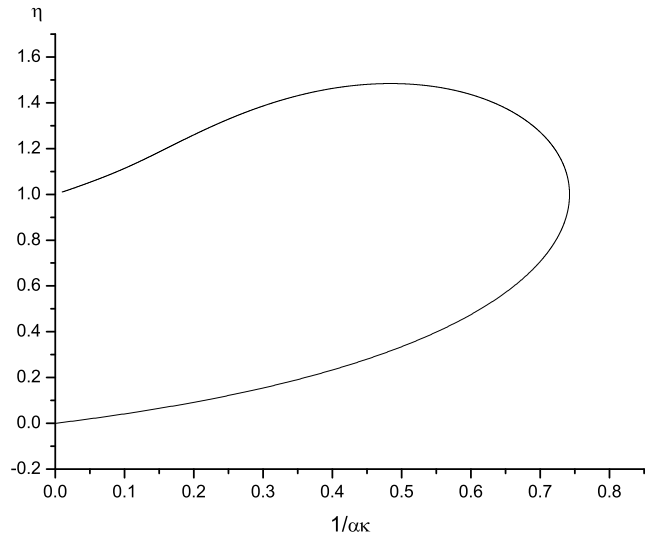}
\end{center}
\caption{Solution of Eq. (\ref{rg8}) for the scaling function $\eta$ as function of $\kappa\alpha$. The branch at $\eta<1$ is not accessible with the initial values of (\ref{bc}). Note that $\kappa=\pi^2$ for the CL system.}
\end{figure}

So far the RG flow was determined in general, without a necessity to identify the relevant differential equation. We proceed to show that the differential equation for $f(\omega)$ determines the function $\eta(\alpha)$ completely and therefore also the flow of ${\bar \alpha}(\omega)$. Consider Eq. (\ref{f5}) that leads to
\begin{equation}\label{rg5}
f''(\omega_c)=\pi\eta\alpha\frac{\sum_na_nn^4}{\pi f(\omega_c)}=\pi\eta\frac{\sum_na_nn^2}{\kappa\omega_c}
\end{equation}
Differentiation of the second equation in (\ref{rg1}) leads to
\begin{eqnarray}\label{rg6}
f''(\omega)&=&\pi(\alpha\eta)'\frac{d{\bar \alpha}}{d\omega}\sum_na_nn^2
\end{eqnarray}
where $(\alpha\eta)'=\frac{d}{d\alpha}[\alpha\eta(\alpha)]$. Equating the last equation at $\omega_c$ with (\ref{rg5}) leads, in terms of $\ell=-\ln\omega$, to
\begin{equation}\label{rg7}
\frac{d\alpha}{d\ell}|_{\omega_c}=-\frac{\eta}{\kappa(\alpha\eta)'}
=-\frac{1}{\kappa}-\frac{1}{\kappa^2\alpha}+O(1/\alpha^2)
\end{equation}
To obtain the expansion we use the perturbative form of $\eta$ in (\ref{bc}). Remarkably, the result (\ref{rg7}) is precisely the two loop RG result For the CL system \cite{hofstetter} (with $g=\pi^2\alpha/2$ in the notations or Ref \onlinecite{hofstetter} and $\kappa=\pi^2$ for the CL system). Note that the same perturbative form in Eq. (\ref{rg2}) yields only the first term $1/\kappa$.

Comparing Eqs. (\ref{rg2},\ref{rg7}) yields
\begin{equation}\label{rg8}
\eta=1+\frac{1}{\kappa
\alpha\eta -1+\kappa\alpha^2\frac{d\eta}{d\alpha}} \,.
\end{equation}
This relation generates a large $\alpha$ expansion with the
leading form $\eta=1+(\kappa \alpha)^{-1}+O(\alpha)^{-2}$,
consistent with the perturbation expansion Eq. (\ref{bc}). It is
remarkable that the perturbation expansion allows for an
asymptotic expansion of (\ref{rg8}), i.e. a different form of
$\eta(\alpha)$ in (\ref{bc}) would not allow such an expansion.

Fig. 1 shows the solution of this equation, with the exact analytic solution given in appendix A.
Note the turning point at $\eta=1,\, 1/(\alpha\kappa)=0.742$.
This corresponds to a fixed point at $\alpha_c$, i.e. if this point is reached at a frequency $\omega_{a}$ then at $\omega<\omega_{a}$ ${\bar\alpha}(\omega)=\alpha_c$ remains constant and
$f(\omega)=\alpha_c\omega$. This behavior is in fact inconsistent with the assumed form (\ref{G'}).
Another difficulty is that continuity of $f'(\omega_0)/f(\omega_0)=\eta[{\bar\alpha}(\omega_0)]=\eta'$ needs $\eta[{\bar\alpha}(\omega_0)]\approx 2$, which is not achieved in Fig. 1.

In the next section we evaluate $f(\omega)$ itself and show that the solution based on Eq. (\ref{f5}) does not satisfy criterion (ii) below some low frequency $\omega_b>\omega_a$. In the latter range one needs to address Eq. (\ref{f4}). we note that the $\omega f''(\omega)$ term in (\ref{f4}) is small at the initial range of $\omega$, e.g. at $\omega_c$ it is $O(1/\alpha)$ relative to the $f'(\omega)$ term. Therefore, to be consistent with the terms neglected due to the criteria (i), we need to start with Eq. (\ref{f5}), and only at the frequency $\sim\omega_b$ we shift to Eq. (\ref{f4}).

We proceed to study the RG form of (\ref{rg4}). Taking a derivative of the second equation in (\ref{rg1}) at $\omega=\omega_c$ yields
\begin{equation}\label{rg9}
C(\alpha)=\alpha[\eta(\alpha)-1](\alpha\eta)'
\end{equation}
Eq. (\ref{f4}) taken at $\omega_c$ yields ${\tilde \eta}(\alpha)=\eta(\alpha)-C(\alpha)/\alpha$. Next we evaluate $f'''(\omega_c)$ in two ways: First, by taking a derivative of Eq. (\ref{f4}) that leads to $f'''(\omega_c)=-\pi{\tilde\eta}(\alpha)\sum_na_nn^2/(\kappa\omega_c^2)$. Second, by taking a derivative of the third equation in (\ref{rg1}). Equating these two forms leads to, finally,
\begin{equation}\label{rg10}
\eta(\alpha)-\frac{C(\alpha)}{\alpha}=\kappa\{C(\alpha)-C'(\alpha)\alpha[\eta(\alpha)-1]\}
\end{equation}
Together with (\ref{rg9}) this is a 2nd order differential equation for $\eta(\alpha)$. We solve this equation by matching at some $\alpha$ to the solution of (\ref{rg8}), as shown in Fig. 2. Curiously, (\ref{rg10}) has an exact solution $\eta=1+1/(\kappa\alpha)$ which gives the one loop solution in (\ref{rg7}). As mentioned above, we apply this solution only below some low frequency $\sim\omega_0$, to be studied in the next section i.e. $\eta=1+\frac{1}{\kappa\alpha}$, does not have then the proper boundary conditions.

\begin{figure}[htb]
\begin{center}
\includegraphics[scale=1.0]{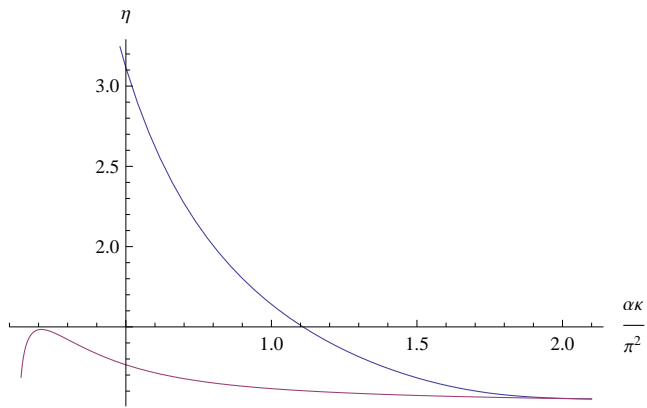}
\end{center}
\caption{Solution of Eq. (\ref{rg10}), upper line,  for the scaling function $\eta$ as function of $\kappa\alpha$. The initial values for this solution are taken from a point on the solution of Eq. (\ref{rg8}), lower line.}
\end{figure}

\section{Solutions for various systems}

We present now explicit solutions for $f(\omega)$ and study the validity criteria. We start with the mathematically simplest case, the CL system.

\subsection{Caldeira Legget system}

Considering Eq. (\ref{f5}) we obtain by differentiating
\begin{equation}\label{s1}
f''(\omega)=\frac{f'(\omega)}{\pi f(\omega)}
\end{equation}
which upon integration yields
\begin{equation}\label{s2}
f'(\omega)=\pi^{-1}\ln [Kf(\omega)]
\end{equation}
where $K$ is an integration constant. Further integration yields
\begin{equation}\label{s3}
li[Kf(\omega)]=\pi^{-1}K(\omega-\omega_a)+li[Kf(\omega_a)]
\end{equation}
where at $\omega_a$, $\bar \alpha(\omega)$ reaches the fixed point of Eq. (\ref{rg8}) i.e. $\bar \alpha(\omega_a)=\alpha_c$, anticipating that this equation is not valid all the way to $\omega_a$;
here $li(x)=\int_0^x\frac{dx'}{\ln x'}$ is the Log Integral
function. $K$ is determined by the $\omega_c$ values
\begin{equation}\label{s4}
f'(\omega_c)=\eta\pi \alpha=\pi^{-1}\ln [Kf(\omega_c)]
\end{equation}
so that
\begin{equation}\label{s5}
Kf(\omega_c)=e^{\eta\pi^2 \alpha}\gg 1
\end{equation}
and from $f(\omega_c)=\pi\alpha\omega_c$ we have
\begin{equation}\label{s8}
K=\frac{e^{\pi^2\alpha\eta}}{\pi\alpha\omega_c} \,.
\end{equation}

Eq. (\ref{s2}) at $\omega=\omega_a$ yields
\begin{equation}\label{s7}
Kf(\omega_a)=e^{\eta'\pi B\omega_a}
\end{equation}
For $B$ and $\omega_a$ we need to solve the coupled equations
\begin{eqnarray}\label{s6}
\int_{e^{\eta'\pi B\omega_a}}^{e^{\eta\pi^2\alpha}}\frac{dz}{\ln
z}&=&\frac{K}{\pi}(\omega_c-\omega_a)\nonumber\\
KB\omega_a^2&=&e^{\eta'\pi B\omega_a}\,.
\end{eqnarray}

An explicit solution requires an asymptotic expansion of $li(x)$,
which is provided by our
RG method. As discussed in section V, the solution has the form $f(\omega)=\omega g(K\omega)$ such that $g(x)$ is $\alpha$ independent. Eqs. (\ref{rg4},\ref{s8}) then yield$f(\omega)=\pi\omega {\bar
\alpha}(\omega)$ where ${\bar \alpha}(\omega)$ the solution of
\begin{equation}\label{alphabar}
K\omega=\frac{e^{\pi^2{\bar \alpha}(\omega)\eta[{\bar
\alpha}(\omega)]}}{\pi {\bar \alpha}(\omega)} \,.
\end{equation}
Inverting this relation we find
\begin{equation}\label{g2}
f(\omega)=\pi\omega{\bar \alpha}(\omega)=\frac{\omega}{\pi\eta}\ln
(\pi{\bar \alpha}(\omega) K\omega)= \frac{\omega}{\pi\eta}\ln
(\frac{K\omega}{\eta\pi}\ln (\pi{\bar \alpha}(\omega)
K\omega))=\frac{\omega}{\pi\eta}\ln (\frac{K\omega}{\eta\pi}\ln
(\frac{K\omega}{\eta\pi}...))
\end{equation}
and at least two $\ln$ embeddings are needed for a large $\alpha$
solution, i.e.
\begin{equation}\label{g3}
f(\omega)=\frac{\omega}{\pi}\ln [\frac{K\omega}{\pi}\ln
\frac{K\omega}{\pi}]+O(\frac{\omega}{\ln (K\omega)})\,.
\end{equation}

The boundary condition at $\omega_a$ is $K\omega_a=e^{\eta'\pi
B\omega_a}/B\omega_a$ so that $g(K\omega_a)=B\omega_a$ becomes
\begin{equation}\label{B2}
g(\frac{e^{\eta'\pi B\omega_a}}{B\omega_a})=B\omega_a\,.
\end{equation}
This equation does not involve the large parameter $\alpha$, hence
 $B\omega_a\approx 1$, $K\omega_a\approx 1$, and the effective mass at scale $\omega_a$ is
\begin{equation}\label{B1}
B\approx \frac{1}{\omega_a}\approx
\frac{e^{\pi^2\alpha}}{\alpha\omega_c}\,.
\end{equation}
We note that Eq. (\ref{alphabar}) implies that ${\bar\alpha}(\omega_a)=O(1)$ i.e. in the vicinity of the fixed point $\alpha_c=0.14$.

We check now the conditions (i)-(iii) for $\omega$ near $\omega_a$. For condition (i) we use
Eq. (\ref{f5}) at $\omega=\omega_a$, where $f'(\omega_a)=\pi\eta[{\bar\alpha}(\omega_a)]{\bar\alpha}(\omega_a)=O(1)$, hence
\begin{equation}\label{c1}
\int_{\omega_a}^{\omega_c}\frac{d\omega_1}{\pi f(\omega_1)}\approx
\ln\alpha
\end{equation}

Hence the condition (i) is satisfied only for $\ln\alpha\gg 1$. The condition (ii) corresponds to
 $\omega f''(\omega)=\omega f'(\omega)/\pi f(\omega)\ll f'(\omega)$, hence $\pi f(\omega)/\omega=\pi^2{\bar\alpha}(\omega)\gg 1$. At $\omega_a$ this condition fails.

 We consider therefore the previous solution as valid  only down to a frequency $\omega_b$, to be determined below. At $\omega<\omega_b$ we use the more complete Eq. (30). As seen in Fig. 2, below $\omega_b$ the slope $\eta(\alpha)$ increases rapidly towards the value $\eta'\approx 2$ which determines
 $\omega_0$. A numerical fit to Fig. 2 and use of (38) yields a weak $\alpha$ dependence, i.e.
 $\omega_0\approx \omega_b/\alpha^x$ with $x\approx 0.2$. Neglecting this effect, we identify $\omega_0\approx \omega_b$ and check the various conditions. Consider first
 \[ \int_{\omega_b}^{\omega_c}\frac{d\omega_1}{f(\omega_1)}=-\pi\ln \frac{f'(\omega_b)}{\pi\alpha} =
 \pi\ln \frac{\alpha}{{\bar\alpha}(\omega_b)\eta[{\bar\alpha}(\omega_b)]}\]
 Since $B\omega_0\approx f(\omega_b)/\omega_b=\pi{\bar\alpha}(\omega_b)$ condition (i) is satisfied for any choice of $\omega_b$ such that ${\bar\alpha}(\omega_b)\gg 1$, e.g. ${\bar\alpha}(\omega_b)=\alpha^{\nu}$ with $\nu<1$.
 \[\omega_0\approx \frac{\alpha\omega_c}{{\bar\alpha}(\omega_b)}\eexp{-\pi^2\alpha+\pi^2{\bar\alpha}(\omega_b)}\]
 with $\eta, \eta({\bar\alpha}(\omega_b))\approx 1$; since $\nu<1$ ${\bar\alpha}(\omega_b)<\alpha$ and $\omega_0\ll\omega_c$ and provides a huge range where Eq. (\ref{f5}) is valid.

 Consider next condition (ii),
 \[\frac{\omega_bf''(\omega_b)}{f'(\omega_b)}=\frac{\omega_b}{\pi f(\omega_b)}=\frac{1}{\pi {\bar\alpha}(\omega_b)} \ll 1\]
 which is also satisfied when ${\bar\alpha}(\omega_b)\gg 1$; condition (iii) is also obvious from
 $B\omega_0\approx \pi{\bar\alpha}(\omega_b)$. Finally we find:
\begin{equation} \label{BB}
B\approx\frac{\pi\alpha^{2\nu}}{\alpha\omega_c}\eexp{\pi^2\alpha-\pi^2\alpha^{\nu}}
\end{equation}

The choice of the exponent $\nu$ is a balance for allowing a maximal range for Eq. (\ref{f5}), which neglects the terms in the 3 conditions on equal footing, and the necessity of satisfying the conditions. We expect then $\nu\ll 1$.
 We note that the result (\ref{BB}) is closer to the Monte Carlo form \cite{lukyanov2} $B\sim\eexp{\pi^2\alpha}/\alpha^2\omega_c$ than (\ref{B1}) above.

\subsection{Study of the general case}
We present here an analysis of the general case, using the asymptotic expansion in the parameter
$\kappa\alpha$. Define the function
\begin{equation}\label{F}
F(x)=\pi \eta \alpha \sum_n a_n n^2e^{-n^2x/\pi}
\end{equation}
so that Eq. (\ref{f5}) becomes
\begin{equation}\label{F1}
f'(\omega)=F[\int_{\omega}^{\omega_c}d\omega_1/f(\omega_1)]\,.
\end{equation}
The boundary condition for this 1st order equation is
$f(\omega_c)=\pi\omega_c\sum_n\alpha_n n^2$ while the condition
$f'(\omega_c)=F(0)$ follows from the equation itself. We now
generate a 2nd order equation
\begin{equation}\label{F2}
\frac{d}{d\omega}F^{-1}[f'(\omega)]=\frac{-1}{f(\omega)}\,.
\end{equation}
Multiplying by $f'(\omega)$ and integrating yields
\begin{equation}\label{F3}
- H[f'(\omega)] + H[f'(\omega_c)]= \int_{\omega}^{\omega_c}f'(\omega_1)\frac{d}{d\omega_1}
F^{-1}[f'(\omega_1)]d\omega_1=-\ln\frac{f(\omega_c)}{f(\omega)}\,.
\end{equation}
Hence in term of the function $H(y) =  \int^{F^{-1}(y)} F(x) dx$, determined up to one integration constant, one
obtains:
\begin{eqnarray}\label{HK}
H[f'(\omega)]&=&-\ln [Kf(\omega)]\nonumber\\
K&=&\frac{e^{-H[f'(\omega_c)]}}{f(\omega_c)}
\end{eqnarray}
For the Caldeira-Legget system one can choose $H[f'(\omega)]=-\pi f'(\omega)$, i.e. a $\alpha$ independent
function, which leads to the solution in the previous section. In the general case however the function
$H(y)$ depends explicitly on $\alpha$, in the form $H(y)=\alpha h(y/\alpha)$ where $h$ is the reciprocal
function of $F/\alpha$. Hence it is not strictly possible to look for a solution of the form $f(\omega)=\omega g(K \omega)$ with a
$\alpha$ independent $g$. Explicit integration of (\ref{HK}) is then required with proper matching (\ref{forms}) at frequency $\omega_0$ but this will not be attempted here in full generality. For $H(y)$ a power law however, one can redefine a scaling function as shown in the next section.

Instead we will follow an approximate method which is consistent with the one loop RG. The idea is to determine the integration constant $K$ by an expansion
near $\omega_c$ where
\[y=F(x)=\pi\eta \alpha \sum_n a_n n^2-\eta \alpha x\sum_n a_n n^4\]
This identifies $F^{-1}(y)$ and the function $H$ is then, to 1st
order in $f'(\omega)-f'(\omega_c)$,
\[H[f'(\omega)]=-\frac{f'(\omega_c) f'(\omega) }{\eta \alpha \sum_n a_n n^4}
\,.\] Using the boundary condition and (\ref{HK})
\begin{equation}\label{K}
K=\frac{\eexp{\eta\alpha\kappa}}{\alpha \pi\omega_c\sum_n a_n
n^2}\,.
\end{equation}

We can now rederive the RG equation (\ref{rg8}) by
 a solution of the form
 $f(\omega)=\omega g(K\omega)$ such that $g(x)$ does not depend
 explicitly on $\alpha$. At $x=K\omega_c$ we have
 \begin{eqnarray}
g(x)&=&\pi\alpha\sum_n a_n n^2 \nonumber\\
xg'(x)&=&\frac{K(\alpha)}{K'(\alpha)} \pi\sum a_n n^2
\end{eqnarray}
so that $f'(\omega_c)=\eta\pi\alpha\sum_n a_n n^2$ becomes eq. (\ref{rg8}).

The reasoning below Eq. (\ref{alphabar}) can now be repeated so that
$f(\omega)$ is generated by repeated $\ln$ embeddings.
 For the effective mass $B$
we need the boundary condition at $\omega_0$, i.e.
$H[f'(\omega_0)]=H(\eta'B\omega_0)=-\ln (KB\omega_0^2)$ and
$g(K\omega_0)=B\omega_0$ which yield an equation for the product
$B\omega_0$,
\begin{equation}\label{Bomega}
g(\frac{e^{-H(\eta'B\omega_0)}}{B\omega_0})=B\omega_0 \,.
\end{equation}
The relation $g(K\omega_0)=B\omega_0$ determines then $\omega_0$
and hence, finally, $B$. Before reaching $\omega_0$, at $\sim\alpha\omega_0$, we expect the modification as discussed in the CL system (previous Section), leading to a change in the exponent $\mu$.

\subsection{Charge Metal system}

In this system we define a mean free path $l$, Fermi wavevector
$k_F$ and then \cite{guinea,golubev} the Fourier expansion is
identified by
\begin{equation}\label{m0}
 [1-\frac{1}{\sqrt{4r^2\sin
^2(z/2)+1}}]=\sum_n a_n(1-\cos nz)/2\,.
\end{equation}
Hence $a_n\approx \frac{2}{\pi r}\ln (r/n)$ for
$1<n\lesssim r$, where $r=R/l$, while
$a_n\approx 0$ otherwise. Applying $d^2/dz^2$ and $d^4/dz^4$ at $z=0$ we get $\sum_n a_n n^2=2  r^2$ and $\sum_n a_n n^4=2  r^2 + 18 r^4$.

We show first that the dependence of the effective mass on the radius is $B\sim R^2$ when $R\rightarrow \infty$, as
for a free particle. We rely here on the proof of section III, within the variational method, that the effective mass can be found from $\phi=0$ in Eq. (\ref{Z1}).  The action has then the form
\[S_{int}\{\theta_0\}=\sum_{n=1}^{n=r}(1/r)\ln(r/n)\bar{S}\{
n\theta_0(\tau)\}\rightarrow -\int_0^1 dx \ln x \bar{S}\{x\bar
{\theta}_0(\tau)\}\]
 where $\bar{S}$ is a functional of $\theta_0(\tau)$, the latter is rescaled
as ${\bar \theta}_0(\tau)=r\theta_0(\tau)$. The action (including
the free term $S_1$) is then r independent and therefore the
effective mass for $[{\bar \theta}_0(\tau)]^2$ is r independent,
which after unscaling yields $B\sim r^2$.

We proceed to study the variational solution at large $r$ and large $\alpha$. While realistic metals have $\alpha\lesssim 1$, this study supplements MC studies \cite{kagalovsky}, done at small $\alpha$. In the region $x \ll \pi$ and in the large $r$ limit, the function $F(x)$ takes the
form:
\begin{eqnarray}
F(x)&=& r^2 \tilde F(r^2 x) \quad , \quad \tilde F(y) \approx 2 \pi \eta \alpha \int_0^1 dz \ln(1/z) z^2 e^{-z^2 y/\pi}
\end{eqnarray}
with $\tilde F(0)=2 \pi \eta \alpha$ and $\tilde F(y) \approx 2\pi^{5/2} \eta\alpha  y^{-3/2}$ for large $y=r^2 x$. For even larger values of $x> \pi$, i.e. $y>r^2 \pi$ the function behaves as in the CL regime $F(x) = \frac{2\pi \eta\alpha e}{r}e^{-x/\pi}$.

It is useful to define the rescaled function via $f(\omega)= r^2 \tilde f(\omega)$ so that in regime $x<\pi$ the variational equation becomes, for $\omega < \omega_c$:
\begin{eqnarray} \label{varnew}
\tilde f'(\omega) = \tilde F( \int_{\omega}^{\omega_c} \frac{d \omega_1}{\tilde f(\omega_1)})
\end{eqnarray}
which is now independent of $r$. It can be solved in principle and assuming that the matching frequency $\omega_0$ occurs in this region $x<\pi$ we get $B=r^2 \tilde B$ with $r$- independent conditions $\tilde f(\omega_0)=\tilde B \omega_0^2$ and $\tilde f'(\omega_0) = \eta' \tilde B \omega_0$ for $\omega_0$ and $\tilde B$. Hence $\omega_0$ and $\tilde B$ are $r$ independent in the large $r$ limit (they depend on $\alpha$) and we recover that $B \sim r^2$.

Note that in the regime of large $y$ we can use the asymptotic form and the equation (\ref{HK}) can then be integrated as:
\begin{equation}\label{H2new}
H[\tilde f'(\omega)]=-2\pi\left(2\pi\eta\alpha \right)^{2/3}[\tilde f'(\omega)]^{1/3}=-\ln
[K \tilde f(\omega)]
\end{equation}
where $K$ is an $\alpha$ dependent integration constant. We note that a scaling function can be defined via
 \[ {\bar f}(\omega)=(2\pi\eta\alpha)^2{\tilde f}(\omega)=\omega g({\bar K}\omega)\]
 where ${\bar K}=K/(2\pi\eta\alpha)^2$. $g(x)$
 satisfies $\pi[g(x)+x g'(x)]=\ln^3 [xg(x)]$, hence $g(x)$ is $\alpha$ independent, except through its argument ${\bar K}$.

 If (\ref{H2new}) is used to identify the matching point (\ref{allbc}) at $\omega_a$ then
 \begin{eqnarray}
&& K{\tilde B}\omega_a^2=\eexp{\eta'{\tilde B} \omega_a(2\pi\eta\alpha)^2/3}\nonumber\\
 && g(\frac{1}{(2\pi\eta\alpha)^2{\tilde B}\omega_a}\eexp{\eta'{\tilde B} \omega_a(2\pi\eta\alpha)^2/3})=
 (2\pi\eta\alpha)^2{\tilde B}\omega_a
 \end{eqnarray}
 so that
 \begin{equation}\label{dmfixed}
 (2\pi\eta\alpha)^2{\tilde B}\omega_a=O(1)
 \end{equation}
 As we show momentarily, this analysis fails near $\omega_a$ as conditions (i), (ii) fail.
As in the CL case, we define $\omega_b$ ($\omega_a<\omega_b\ll \omega_c$) so that below $\omega_b$ the corrected Eq. (\ref{f4}) is applied and then we expect $\omega_0\approx\omega_b$.

 Alternatively, we can use a scaling form for ${\bar f}(\omega)$ as in Eq. (\ref{rg1}) i.e.
  ${\bar f}(\omega)=2\pi\omega{\bar \alpha}(\omega)\, , {\bar f}'(\omega)=2\pi\omega{\bar \alpha}(\omega)\eta[{\bar \alpha}(\omega)]$ . Hence (\ref{H2new}) becomes
\begin{equation} \label{BB3}
{\bar K}\omega=\frac{1}{\pi{\bar \alpha}(\omega)}\eexp{[\pi^2{\bar \alpha}(\omega)\eta({\bar \alpha}(\omega))]^{1/3}}
\end{equation}
which, at $\omega=\omega_c$, identifies $K=\eexp{2\pi^2\eta\alpha}/(2\pi\alpha\omega_c)$.
Matching at $\omega_0\approx \omega_b$ and using (\ref{varnew})
\begin{equation}
 \frac{1}{{\tilde B}\omega_0}\approx \frac{(2\pi\eta\alpha)^2}{2\pi{\bar \alpha}(\omega_b)}
 \end{equation}
 Note that replacing $\omega_b\rightarrow \omega_a$ recovers Eq. (\ref{dmfixed}), confirming that  ${\bar\alpha}(\omega_a)=O(1)$.

 For condition (i) we need
 \begin{equation}
 \int_{\omega_b}^{\omega_c}\frac{d\omega_1}{{\tilde f}(\omega_1)}=\frac{\pi(2\pi\eta\alpha)^2}{[2\pi{\bar \alpha}(\omega)\eta({\bar \alpha}(\omega))]^{2/3}}\gg\frac{1}{{\tilde B}\omega_0}
 \end{equation}
  which is satisfied if ${\bar \alpha}(\omega_b)\gg 1$; clearly at $\omega_a$ this condition fails. For condition (ii), by a derivative of Eq. (\ref{varnew}), we obtain
 \[ \frac{\omega_b f''(\omega_b)}{f'(\omega_b)}=\frac{3}{2}\frac{[\pi{\bar \alpha}(\omega_b)\eta({\bar \alpha}(\omega_b))]^{2/3}}{2\pi^2 {\bar \alpha}(\omega_b)}\sim [{\bar \alpha}(\omega_b)]^{-1/2}\ll 1\]
 which is also satisfied when ${\bar \alpha}(\omega)\ll 1$. Finally, condition (iii) is satisfied since $B\sim r^2$.

 To obtain the effective mass $B$, Eq. (\ref{BB3}) leads to
 \[ \omega_b= \frac{\alpha\omega_c(2\pi\eta\alpha)^2}{{\bar \alpha}(\omega_b)}\,
 \eexp{-2\pi^2\eta\alpha+[2\pi^4{\bar \alpha}(\omega_b)]^{1/3}}\]
 where $K$ from (69) is used and $\kappa=2\pi^2/9$ at $r\gg 1$. As in the CL case, we choose ${\bar \alpha}(\omega)\sim\alpha^{\nu}$ with $\nu\ll 1$ so that $\omega_b\ll \omega_c$, providing a large integration regime for Eq. (32). Finally, the effective mass is
 \begin{equation}\label{BBB}
 {\tilde B}=\frac{2\pi {\bar \alpha}^2(\omega_b)}{(2\pi\eta\alpha)^4\alpha\omega_c}\eexp{2\pi^2\eta\alpha-
 [\pi^2{\bar \alpha}(\omega_b)]^{1/3}}\approx \frac{1}{(2\pi)^3\alpha^5\omega_c}\eexp{2\pi^2\alpha}
\end{equation}

It is easy to see that the condition that the frequency $\omega_0$ belongs to the scale invariant regime and not in the CL regime is $x \ll \pi$ i.e.:
\begin{eqnarray} \label{varnew2}
\pi r^2 \gg  \int_{\omega_0}^{\omega_c} \frac{d \omega}{\tilde f(\omega)}
\end{eqnarray}
the r.h.s. which can be determined from the solution, depends only on $\alpha$ and not on $r$, hence this sets a minimum radius as a condition.

\section{Correlation function}

\subsection{small $\alpha$ perturbation theory}

Independently of the variational method it is also useful to consider the straight small $\alpha$ perturbation theory of the action (\ref{Z1}). We consider first the effect of the $\phi$ integration in Eq. (\ref{Z1}). Perturbation expansion in $\alpha$ leads in general to a $\phi$ dependence of the form
$\eexp{2\pi i\phi{\tilde \tau}/\beta}$ where ${\tilde \tau}$ is a linear combination of the various time variables $\tau_i$ in the expansion. The $\phi$ integral is then
\begin{equation}\label{Ksum}
\sum_K\int_{-\infty}^{\infty}d\phi \eexp{2\pi
i\phi(K+\phi_x)-\frac{2\pi^2MR^2\phi^2}{\beta}+2\pi i\phi{\tilde \tau}/\beta}=
\sum_K \eexp{-\beta (K+\phi_x+{\tilde \tau}/\beta)^2/(2MR^2)}
\end{equation}
 We expect that the various $\tau_i$ integrations converge so that at $\beta\rightarrow\infty$ the limit ${\tilde \tau}/\beta\ll K+\phi_x$ can be taken and then the sum is dominated by $K=0$ when $|\phi_x|<\half$. We show this explicitely for the 1st order below.

 For $\phi_x=0$ and $\beta=\infty$ of (\ref{Z1}) we can therefore consider $Z_{\phi=0}$. Here we compute the correlation function of $\cos \theta$ order by order in $\alpha$. The zero-th order is obtained from the free particle action $S_1$ (\ref{S}) and given by:
\begin{eqnarray}
\langle \cos(\theta(\tau)-\theta(0)) \rangle_0 = \exp( - \int_{-\infty}^{+\infty} \frac{d \omega}{2 \pi} (1-\cos(\omega \tau)) \frac{1}{M R^2 \omega^2} ) = \exp(- \frac{1}{2} \omega_M |\tau|)
\end{eqnarray}
where we have defined $\omega_M=\frac{1}{M R^2}$. To perform the expansion we take the $\beta\rightarrow \infty$ limit in the time integrals since these are found to be convergent, while we keep $\beta$ in the $\phi$ dependence. For $\phi_x=0$ we will rewrite the interaction in Eq. (\ref{Z1}):

\begin{eqnarray}
&& S_{int}=  - \frac{1}{2} \alpha \sum_{n \geq 1} a_n \int_{-\infty}^{\infty}\int_{-\infty}^{+ \infty} d\tau d\tau '
\frac{ \cos \{n[\theta(\tau)-\theta(\tau')+2\pi\phi(\tau_1-\tau_2)/\beta]\}-1}{|\tau -\tau '|^2}   \label{newsint}
\end{eqnarray}

The first order correction is obtained from the connected average, using Eq. (\ref{Ksum}):
\begin{eqnarray}
&& \langle \cos(\theta(\tau)-\theta(0)) \rangle_1 = \frac{\alpha}{2}
\sum_{n \geq 1} a_n\int_{\tau_1}\int_{\tau_2}S_K(\tau,\tau_-\tau_2)\frac{1}{(\tau_1-\tau_2)^2}
\langle \eexp{i[\theta(\tau)-\theta(0)+n\theta(\tau_1)-n\theta(\tau_2)]}\rangle_{0,c} \nonumber\\
&& = \frac{\alpha}{2} a_1 \int_{\tau_1}\int_{\tau_2}S_K(\tau,\tau_1-\tau_2)\frac{e^{- \frac{\omega_M}{2} ( |\tau|+|\tau_1-\tau_2|)}}{(\tau_1-\tau_2)^2}
( e^{\frac{\omega_M}{2} ( |\tau-\tau_1| + |\tau_2|  -|\tau_1|-|\tau-\tau_2|)} - 1)\nonumber
\end{eqnarray}
where $S_K(\tau,x)=\sum_K \eexp{-\beta (K+ (\tau+x)/\beta)^2/(2MR^2)}$. As we find the integrals are indeed convergent so that $\beta\rightarrow \infty$ can be taken and $S_K(\tau,x)\rightarrow 1$. We have discarded exponentially decaying terms in $\tau$ such as produced by $n >1$.
It is important to note that the starting integral is convergent for $\tau_1 \approx \tau_2$. To see that one can symmetrize in $\tau_1,\tau_2$ the term in parenthesis: expansion for $\tau_1\approx \tau_2$ then yields an additional $(\tau_1-\tau_2)^2$ term. This is a general property for all connected averages: the small time apparent singularity is absent. Indeed, expanding the cosine in the vertex (\ref{newsint}) and contracting the two fields with times external to the vertex yields $\nabla \theta(\tau_1) \theta(\tau)$ correlations, which are always bounded in the action $S_1$.

Although integral (\ref{integr}) is tedious to compute, its large $\tau$ behaviour is easily extracted. It is clear that, assuming a Wick decoupling for the $\eexp{i\theta}$ factors, then the exponential decay of each 2-point correlator fixes the value of $\theta_1\rightarrow 0$ and of $\theta_2\rightarrow \tau$, leading to a $1/\tau^2$ form. To be more precise, the mass term forces the variables $\tau_1 = 0 + O(1/\omega_M)$ and $\tau_2 = \tau + O(1/\omega_M)$, i.e. this is the region which dominates the integral in (\ref{integr}) at large $\tau \gg 1/\omega_M$. Integration is then easy in that region and amounts to replace $\exp(- \frac{1}{2} \omega_M |\tau_1|) \to \frac{4}{\omega_M} \delta(\tau_1)$ and $\exp(- \frac{1}{2} \omega_M |\tau_2-\tau|) \to \frac{4}{\omega_M} \delta(\tau_2-\tau)$. Note that since the result is only a function of $\omega_M \tau$ the large $\tau$ limit is the same as the large $\omega_M$ limit at fixed $\tau$. This yields:
\begin{eqnarray} \label{firstorder}
&& \langle \cos(\theta(\tau)-\theta(0)) \rangle_1
 \approx \alpha a_1\frac{8}{\omega_M^2 \tau^2}
\end{eqnarray}
at large $\tau$. The amplitude is confirmed by the detailed calculation of the integral given in Appendix \ref{int}, as well as by
a numerical check.

We note that this $1/\tau^2$ decay is in general agreement with the constraints derived in Ref. \cite{spohn} for the long range $XY$ model, very similar to our CL model I. There it was shown that for strictly ferromagnetic LR interactions the spin correlation cannot decay slower than the interaction. For the DM model the $1/\tau^2$ has a $a_1\sim 1/r$ coefficient, hence it vanishes in the $r\rightarrow \infty$ limit.

The second order correction can be written as:
\begin{eqnarray} \label{secondorder}
&&\langle \cos(\theta(\tau)-\theta(0)) \rangle_2 = \frac{\alpha^2}{8} \sum_{n \geq 1} a_{n }\sum_{n' \geq 1}a_{n'}\int_{\tau_1}...\int_{\tau_4}
\frac{\langle\eexp{i[\theta(\tau)-\theta(0)+n\theta(\tau_1)-n\theta(\tau_2)
+n'\theta(\tau_3)-n'\theta(\tau_4)]}\rangle_c}{(\tau_1-\tau_2)^2 (\tau_3-\tau_4)^2 }
\end{eqnarray}
At large $\tau$ the main contribution comes from $\tau_2 \approx \tau$, $\tau_3 \approx 0$ and $\tau_1 \approx \tau_4$ (and one deduced by exchange $1,2$ with $3,4$) and yields:
\begin{eqnarray}
&& \langle \cos(\theta(\tau)-\theta(0)) \rangle_2 \approx 8 \alpha^2
 a_1^2\frac{1}{\omega_M^3} \int_{\tau_1}\frac{1}{(\tau_1-\tau)^2\tau_1^2}
\end{eqnarray}
This integral looks divergent at small times but it is meant to be regularized for $\tau_1$ near zero by the region $\tau_2 \approx \tau$, $\tau_3 \approx \tau_1 \approx \tau_4 \approx 0$ in the above integral (\ref{secondorder}). This mainly replaces the $1/\tau_1^2$ factor by a $\sim \langle \cos(\theta(\tau_1)-\theta(0)) \rangle_1$ factor regular at $\tau_1=0$. Similarly, the singularity at $\tau_1=\tau$ is smoothed by proper integration of (\ref{secondorder}) in the region $\tau_2 \approx \tau_1 \approx \tau_4 \approx \tau$, $\tau_3 \approx 0$. Since it is regularized at small times on times of order $\sim 1/\omega_M$ the above integral behaves as:
\begin{eqnarray}
&& \langle \cos(\theta(\tau)-\theta(0)) \rangle_2 \approx 8 \alpha^2
 a_1^2  \frac{A}{\omega_M^2 \tau^2}
\end{eqnarray}
To compute the coefficient $A$ we need to perform carefully the integrals in the small time regularization region. The question of universality of this amplitude is discussed in Appendix \ref{higherorder}. We will not attempt that here but simply note that there is no large time divergence in the above integral, i.e. the coefficient $A$ is finite and does not contain any log-divergence.

\subsection{large $\alpha$ behaviour via matching}

Let us now estimate the correlation function in the large $\alpha$ limit and we restrict to CL for simplicity. For $\tau$ not
too large we can just use the straight large $\alpha$ perturbation theory:
\begin{eqnarray}\label{match1}
\langle \cos(\theta(\tau)-\theta(0)) \rangle =
\exp( - \int_{-\omega_c}^{\omega_c} \frac{d \omega}{2 \pi} (1-\cos(\omega \tau)) \frac{1}{\alpha |\omega|} ) \approx \frac{1}{(\omega_c \tau)^{\frac{1}{\pi \alpha}}}
\end{eqnarray}
which presumably is valid only for $\ln (\omega_c \tau) \ll \pi \alpha$. For larger time one needs
to consider renormalization of the dissipation. We use the analysis of the previous sections.
For large $\alpha$ we expect that we can use the fixed point action which is of the form (\ref{S})
with renormalized parameters, i.e. near $\alpha_c=1/(0.742\kappa)$ and with the mass $M$ replaced by the renormalized mass $B/R^2$ with $B=B(\alpha)$. To get an estimate of the correlation function at large
$\tau$, we can now use the above result (\ref{firstorder}) for the small coupling expansion replacing $\alpha$ by $\alpha^*$ and $\omega_M$ by $1/B(\alpha)$ (according to Eq. (\ref{BB})).  Since $\alpha^*$ is not strictly small this will only provide an estimate. One gets, keeping the dominant exponential term,
\begin{eqnarray}\label{match2}
\langle \cos(\theta(\tau)-\theta(0)) \rangle \sim \frac{e^{2 \pi^2 \alpha}}{(\omega_c \tau)^2 }
\end{eqnarray}
which we expect to be valid for $\ln(\omega_c \tau) \gg \pi^2 \alpha + O(\ln \alpha)$. Eq. (\ref{match2}) matches (\ref{match1}) at $\ln \omega_c\tau\approx \pi^2\alpha$.

\section{Discussion}

We have studied two types of environments: (i) The Caldeira Legget (CL) system, with relevance to small rings, $r<1$, or to the Coulomb box problem, and (ii) the dirty metal environment, that can couple to either a charge (CM system) or an electric dipole (DM system), with relevance to experiments on cold Rydberg atoms \cite{hyafil}.

For the CL system, the variational method was shown to be equivalent to an RG scheme, reproducing the known two loop result \cite{hofstetter}. Our method provides an expansion to all orders in $1/\alpha$ and leads to the renormalized mass Eq. (\ref{BB}), which is close to the result of the boundary field theory \cite{lukyanov1} and the MC data \cite{lukyanov2}.

At small $\alpha$ we find a regular expansion without $\ln$ divergences up to second order, i.e. the RG $\beta$-function for $\alpha$ seems to vanish perturbatively. Since we know that the RG flow of $\alpha$ at large $\alpha$ is towards smaller values of $\alpha$, there seems to be three main possible scenarios: (i) the flow towards small $\alpha$ becomes much slower, either exponentially due to some putative non perturbative corrections, or to some higher order in $\alpha$ (ii) there is a line of fixed points for $\alpha < \alpha_c$ with some termination
point $\alpha_c$ (iii) there is a infinite set of fixed points at small alpha with accumulation at zero \cite{new}.

In fact the result $\langle \cos[\theta(\tau)]\cos[\theta(0)]\rangle\sim 1/\tau^2$ is a robust one, relating to a theorem on an XY model on a lattice \cite{spohn}. That this result is derived in first order in $\alpha$ is remarkable. For large $\alpha$ one should use the scaling to small $\alpha$ and then use the former result.

For the dirty metal problem we show that at large $r$ the whole action scales with $r^2$, leading an $r$ independent effective mass $B/R^2$. Furthermore, we find a scaling form for large $r$ and large $\alpha$ that leads to the renormalized mass, Eq. (\ref{BBB}). For the CM system $\alpha\lesssim 1$ from Eq. (\ref{e05}), yet for the DM system a large $\alpha$ may be realized in Eq. (\ref{e010}) if the dipole has $p>e\ell$, i.e. the extension of the Rydberg atom needs to be $>\ell$. The large $\alpha$ solution is useful also as a complement to the small $\alpha$ MC data at $\alpha=0.19$, showing saturation of the effective mass with $r$. Therefore, the claims for an $r$ dependent mass \cite{golubev} are in contrast with both weak and strong $\alpha$ results. We note also that the result to first order $\langle \cos[\theta(\tau)]\cos[\theta(0)]\rangle\sim a_1/\tau^2$ vanishes as $a_1\sim \frac{1}{r}\rightarrow 0$ suggesting that the nonlinearities associated with $\alpha$ become weaker in the large $r$ limit.

We believe that the correspondence of the variational method with the scaling forms is a useful and instructive guide for studies of large variety of nonlinear systems.

\bigskip

We thank I. S. Burmistrov, A. Golub, P. Guinea, V. Kagalovsky, A. D. Zaikin and G. Zarand for stimulating discussions. BH acknowledges kind hospitality and financial support from LPTENS and PLD from Ben Gurion University. This research was supported by THE ISRAEL SCIENCE FOUNDATION (grant No. 1078/07) and by the ANR grant 09-BLAN-0097-01/2.

\newpage

\appendix

\section{The $\eta$ parameter}

We solve here Eq. (\ref{rg8}) for $\eta(\alpha)$. We change variable to $x=1/\alpha$ and then to $y(x)=\eta(x)/x$
\beq{104}
\eta'(x)&=&-\frac{\eta(x)}{\kappa[\eta(x)-1]}+\frac{\eta(x)}{x}\\
y'(x)&=&-\frac{y(x)}{\kappa[xy(x)-1]}\eeq
therefore
\beq{105}-\kappa\frac{xy(x)-1}{y(x)}&=&\frac{1}{y'(x)}=x'(y)\\
x'(y)+\kappa x(y)&=&\frac{\kappa}{y}\eeq
A general solution of the homogenous part is $x=C_1\eexp{-\kappa
y}$ while for a solution to the full equation substitute
$x=A(y)\eexp{-\kappa y}$ so that $A'(y)=\kappa\eexp{\kappa y}/y$,
hence
\beq{106} A(y)&=&\kappa\int\frac{\eexp{\kappa y}}{y}dy=\kappa
\mbox{Ei}(\kappa y)+C_2\\
x(y)&=& (C_1+C_2)\eexp{-\kappa y}+\kappa \mbox{Ei}(\kappa
y)\eexp{-\kappa y}\eeq
where $\mbox{Ei}$ is the exponential integral function, with the
asymptotic expansion
\beq{107}
\mbox{Ei}(z)=\frac{\eexp{z}}{z}[1+\frac{1}{z}+\frac{2!}{z^2}+...]
\qquad z\rightarrow \infty\eeq
The boundary condition gives at $x\rightarrow 0$
\beq{108} x=\frac{1}{y}+\frac{1}{\kappa y^2}+C\eexp{-\kappa y}+...\\
y=\frac{1}{x}+\frac{1}{\kappa}+\frac{C}{x^2}\eexp{-\kappa/x}+...\\
\eta(x)=xy(x)=1+\frac{x}{\kappa}+\frac{C}{x}\eexp{-\kappa /x}\eeq
where $C=C_1+C_2$. Comparison with Eq. (\ref{e03}) shows that
$C=0$, a remarkable result. The solution is then
\beq{109} 1=\kappa\alpha\mbox{Ei}(\kappa\alpha
\eta)\eexp{-\kappa\alpha \eta}\eeq
This result is plotted in Fig. 1 for $\eta$ as function
of $1/(\kappa\alpha)$.

\section{The Log integral}
\label{app:log}

We present here the mathematical result, i.e. solving the Log
integral by using RG and deriving an asymptotic
solution. Consider the equation:
\begin{equation}\label{log1}
\tilde f'(y)=\frac{1}{\pi} \ln [\tilde K \tilde f(y)]
\end{equation}
with the boundary condition $\tilde f(1)=\pi \alpha$. This is the equation for the CL system () in the text, defining
$f(\omega)=\omega_c \tilde f(\omega/\omega_c)$. The constant $\tilde K$ is parameterized as $$\tilde K=e^{\pi^2 \alpha\eta}/\pi \alpha$$
so that $\tilde f'(1)=\pi \alpha\eta$. In general we expect that any pair
$(\alpha,\eta)$ will produce a solution $\tilde f(y)$.

We rewrite the solution in the form
\begin{equation}\label{log2}
\tilde f(y)=yg\left (\frac{e^{\pi^2 \alpha\eta}}{\pi \alpha}y\right )
\end{equation}
so that the boundary condition at $y=1$ is
\begin{equation}\label{log3}
g[\tilde K(\alpha)]=\pi \alpha \,.
\end{equation}
Imagine now a varying boundary condition $\alpha$ and that the
function $g$ does not depend explicitly on $\alpha$; this implies
that $\eta(\alpha)$ must be chosen in a specific way. Eq.
(\ref{log1}) can be written as
\begin{equation}\label{log4}
g(x)+xg'(x)=\frac{1}{\pi} \ln[xg(x)]
\end{equation}
where $x=\tilde Ky$. Taking a derivative of (\ref{log3}) yields
$(d\tilde K/d\alpha)g'(\tilde K)=\pi$ so that Eq. (\ref{log4}) at $x=\tilde K$ yields
\begin{equation}\label{log5}
\alpha+\frac{\tilde K(\alpha)}{\tilde K'(\alpha)}=\frac{1}{\pi^2} \ln (\tilde K(\alpha) \pi \alpha)=\alpha\eta(\alpha)
\end{equation}
leading to a differential equation for $\eta(\alpha)$
\begin{equation}\label{log6}
\eta=1+\frac{\tilde K(\alpha)}{\alpha \tilde K'(\alpha)}=1+\frac{1}{\pi^2 \alpha\eta
-1+ \pi^2 \alpha^2\frac{d\eta}{d\alpha}} \,.
\end{equation}
Integrating this equation from any initial values $(\alpha,\eta)$
yields a function $\eta(\alpha)$; the full solution at $y<1$ is
then obtained as $\tilde f(y) = \pi y \bar \alpha(y)$ where the
function $\bar \alpha(y)$ is determined by the choice $\tilde K[\bar{\alpha}(y)]=\tilde K(\alpha)y$. Indeed one then has:
\[f(y)=yg(\tilde Ky)=yg[\tilde K(\bar{\alpha}(y))]=\pi y\bar{\alpha}(y).\]
Therefore one needs to invert the algebraic relation
$\tilde K[\bar{\alpha}(y)]=\tilde K(\alpha)y$ to find $\tilde f(y)$, leading to a form
like (\ref{g2}). In general, however, (\ref{log6}) is is not
easier than the original (\ref{log1}), except for the initial
values $(\alpha=\infty,\eta=1)$ which allow an asymptotic
expansion with large $\alpha$. That our physical system satisfies
this special tuning is most remarkable.

\section{integration of RG}
It is instructive to study the one loop RG of the dirty metal system,
and compare with the variational solution.

Consider then the RG equations \cite{guinea}, which can also be read off from Eq. (\ref{Gpert2})
\begin{equation}\label{RG1}
\frac{d\alpha_n}{d\ell}=-\frac{n^2\alpha_n}{\pi^2\sum_n\alpha_n
n^2}\,.
\end{equation}
with have defined $\alpha_n=\alpha a_n$. Let us consider the function $F(x)$ of Eq. (\ref{F}) setting $\eta=1$ there.
It becomes now
$\ell$ dependent with:
\begin{equation}\label{RG2}
\frac{dF_{\ell}(x)}{d\ell}=\frac{F'_{\ell}(x)}{ F_{\ell}(0)}
\end{equation}
Change to the variable
\begin{equation}\label{RG3}
F_{\ell}(0)\frac{\partial}{\partial
\ell}=\frac{\partial}{\partial \mu}
\end{equation}
so that
\begin{equation}\label{RG4}
[\frac{\partial}{\partial \mu}-\frac{\partial}{\partial
x}]F_{\ell}(x)=0
\end{equation}
which has the solution
\begin{equation}\label{RG5}
F_{\ell(\mu)}(x)=F_{\ell(0)}(x+\mu(\ell))
\end{equation}
and with $\mu=\int_0^{\ell}d\ell/\pi F_{\ell}(0)$ we have the
general solution
\begin{equation}\label{RG6}
F_{\ell}(0)=F_0(\int_0^{\ell}\frac{d\ell'}{ F_{\ell'}(0)})
\end{equation}
Note that it has some formal similarity to the variational Eq. (\ref{F1}) if we
define $\ell=-\ln \omega/\omega_c$. We proceed to study the dirty metal case, with
$F_{\ell}=F_{\ell}(0)$,
\begin{equation}\label{RG7}
F_{\ell}=\frac{2 \pi^{5/2} \alpha}{r}[\int_0^{\ell}\frac{d\ell'}{
F_{\ell'}}]^{-3/2}
\end{equation}
Hence by differenciating
\begin{equation}\label{RG8}
-\frac{2}{3}(\frac{2\pi^{5/2} \alpha}{r})^{2/3}\frac{\partial
F_{\ell}/\partial \ell}{F^{2/3}_\ell}=1
\end{equation}
Therefore
\begin{equation}\label{RG9}
2 (\frac{2\pi^{5/2} \alpha}{r})^{2/3} [F^{1/3}_{\ell=0}-F^{1/3}_{\ell}]=\ell=\ln(\frac{\omega_c}{\omega_c^R})
\end{equation}
where $\omega_c^R$ is a renormalized cutoff. RG terminates at
$F_{\ell}=1\ll F_0=2 \pi \alpha r^2$, hence
\begin{equation}\label{RG10}
2  (\frac{2\pi^{5/2} \alpha}{r})^{2/3} (2 \pi \alpha
r^2)^{1/3}=\ln(\frac{\omega_c}{\omega_c^R})\,.
\end{equation}
Since powers of $r$ cancel $\omega_c\sim\omega_c^R$, i.e. the frequency at which
the RG is stopped is independent of $r$, a conclusion also obtained in the text.

\section{calculation of an integral} \label{int}

The integral given in the text, upon rescaling $\omega_M \tau \to \tau$ is computed as:
\begin{eqnarray}
&&  \frac{1}{2} \int_{\tau_1}\int_{\tau_2} \frac{e^{- \frac{1}{2} ( |\tau|+|\tau_1-\tau_2|)}}{(\tau_1-\tau_2)^2}
( e^{\frac{\omega_M}{2} ( |\tau-\tau_1| + |\tau_2|  -|\tau_1|-|\tau-\tau_2|)} - 1)\nonumber \\
&&=\half \eexp{-|\tau|/2}\{\int_{-\infty}^{-\tau/2}dx\,  \frac{\eexp{x}}{x^2}[(1-x-\tau/2)(\eexp{\tau}-1)
-\tau]+\int_{-\tau/2}^0 dx \, \frac{\eexp{x}}{x^2}[\eexp{-2x}(1+\tau/2+x)+x-\tau/2-1]\nonumber\\
&&+\int_0^{\tau/2}dx \, \frac{\eexp{-x}}{x^2}[(1-\eexp{-2x})(1-\tau/2+x)-2x]+
\int_{\tau/2}^{\infty}dx \, \frac{\eexp{-x}}{x^2}[(1-\eexp{-\tau})(1-x+\tau/2)-\tau]\}\nonumber\\
&& =  \frac{1}{4} e^{-2 \tau} \bigg(-e^{3 \tau/2} (3 \tau-4) \text{Ei}\left(-\frac{3
   \tau}{2}\right)-e^{\tau/2} \left(e^{2 \tau} \tau+\tau+4\right)
   \text{Ei}\left(-\frac{\tau}{2}\right) \\
   && +e^{3 \tau/2} \left(\tau
   \left(\text{Ei}\left(\frac{\tau}{2}\right)+\log (27)\right)+8-\log
   (81)\right)-4 e^{2 \tau}-4\bigg) \nonumber \\
   && = \frac{1}{4} \tau^2 (-2 \log (\tau)-2 \gamma +1+\log (4))+O\left(t^3\right)  \quad , \quad \tau \ll 1 \\
   && = \frac{8}{\tau^2} + \frac{384}{\tau^4} + O(\tau e^{-\tau/2})  \quad , \quad \tau \gg 1
\label{integr}
\end{eqnarray}

\section{structure of higher orders in small $\alpha$ perturbation} \label{higherorder}

The discussion of the first and second order corrections in the text suggested
that the large time behaviour of the integrals could be obtained from a Wick theorem on the $e^{i \theta}$ fields with a $\delta$ function correlator, e.g. the structure of the second order correction (\ref{secondorder}) at large time is an integral dominated by
$\tau_2 \approx \tau$, $\tau_3 \approx 0$ and $\tau_1 \approx \tau_4$, i.e. by the region such that the
charges in $e^{i n \theta(\tau_i)}$ should compensate. This was found to generically lead to $1/\tau^2$ decay. A question is then whether the amplitude of this decay can be obtained to all orders using this property, and how does it depend on the details of the short time cutoff.

Let us first consider a toy model with the same property. It is a gaussian theory of partition sum $\int Dz(\tau) e^{-S}$ (for $a_n=\delta_{n1}$):
\begin{eqnarray} \label{equiv0}
&& S = \frac{ \omega_M}{4} \int_{\tau} z(\tau) z^*(\tau) - \frac{1}{2} \alpha \int_{\tau,\tau'} z(\tau) g(\tau-\tau') z^*(\tau')
\end{eqnarray}
where $z(\tau)$ plays the role of $e^{i \theta(\tau)}$ and $g(\tau) \approx 1/\tau^2$ at large $\tau$. Being a gaussian variable it reproduces, for $\alpha=0$ the propagator $<z(\tau) z^*(\tau')>_0 = 4 \omega_M^{-1} \delta(\tau-\tau')$. There is thus some similarities in the perturbation expansion in $\alpha$ with the original model. Here however it is immediate to obtain:
\begin{eqnarray}  \label{corr0}
&& <z_\tau z^*_{\tau'}> = G(\tau-\tau') \quad , \quad G(\omega) = \frac4{ \omega_M - 2 \alpha g(\omega)}
\end{eqnarray}
Let us consider two examples for the short time cutoff function in (\ref{equiv0}). (i) $g(\omega)= \pi \omega_M e^{-|\omega|/\omega_M}$ corresponds to a Lorentzian $g(\tau)=1/(\omega_M^{-2} + \tau^2)$. The coefficient of $G(\tau) \sim A/\tau^2$ at large $\tau$ is obtained from the expansion $G(\omega)=G(0)- \pi A |\omega| + ..$ and reads $A=8 \alpha/((1-2 \alpha \pi) \omega_M^2)$ by expanding (\ref{corr0}) (ii) $g(\omega)= - \pi |\omega| e^{-\omega^2/\omega_M^2}$ which gives
$A=8 \alpha/\omega_M^2$, i.e. only a first order contribution, all higher orders being zero.

The above example shows that the amplitude $A$ can depend on the short time cutoff beyond leading order. In that case
it was however easily calculable by convolutions. To check whether one can indeed predict the $1/\tau^2$ coefficient more generally, let us consider now the following general discrete XY model of partition sum:
\begin{eqnarray} \label{equiv}
&& Z = \prod_{i} \int_0^{2 \pi} \frac{d\theta_i}{2 \pi} \exp( \sum_{k,l} g_{kl} e^{i (\theta_k - \theta_l)} )
\end{eqnarray}
setting $g_{kk}=0$ for convenience. A calculation using mathematica then gives, for $x \neq 0$ (here $x,p,q$ belong to an arbitrary lattice):
\begin{eqnarray} \label{equiv1}
&& < e^{i ( \theta_x-\theta_0)} > = g_{0x} + \sum_p g_{0p} g_{px} + \sum_{p,q \neq 0,x} g_{0p} g_{pq} g_{qx}
- \frac{1}{2} g_{x0} g_{0x}^2 \\
&& + \sum_{p \neq q \neq r \neq p \neq 0,x} g_{xp} g_{pq} g_{qr} g_{rx} - \sum_{p \neq 0,x} g_{x0} g_{0x} g_{0p} g_{px} - \frac{1}{2} \sum_{p \neq 0,x} ( g_{0p} g_{xp} g_{px}^2 + g_{p0} g_{px} g_{0p}^2 + g_{0x}^2 g_{xp} g_{0p} ) + O(g^5)
\end{eqnarray}

Taking as an example a one dimensional chain with discrete $\tau$ values, and $g_{\tau \tau'} = g(\tau-\tau') \sim 1/(\tau-\tau')^2$ at large $\tau-\tau'$, one sees that up to order $O(g^3)$ (included) the structure captured by model (\ref{equiv0}) is correct to predict the coefficient of the $1/\tau^2$ decay of $< e^{i ( \theta_\tau-\theta_0)} >$. Indeed, up to that order, the $1/\tau^2$ decay can be obtained from the convolution of the $g$ kernel. However some new terms arise at order $g^4$ which are not of the above form and do contribute to the $1/\tau^2$ decay, and the calculation of $A$ becomes
more complicated than in model (\ref{equiv0}).

\end{document}